\documentclass{ws-procs9x6}

\usepackage{latexsym} 
\usepackage[errorstop]{feynmf}

\def\ket#1{|#1\rangle}
\newcommand{\gsim}{\mathrel{\rlap{\raisebox{.3ex}{$>$}}
    \raisebox{-.6ex}{$\sim$}}}
\newcommand{\lsim}{\mathrel{\rlap{\raisebox{.3ex}{$<$}}
    \raisebox{-.6ex}{$\sim$}}}

\newcommand{\nn}{\nonumber}

\newcommand{\tetaot}{\mbox{$\theta_{13}$}}
\newcommand{\tetatt}{\mbox{$\theta_{23}$}}

\newcommand{\delot}{\mbox{$\Delta_{13}$}}

\newcommand{\bea}{\begin{eqnarray}}
\newcommand{\eea}{\end{eqnarray}}
\newcommand{\be}{\begin{equation}}
\newcommand{\ee}{\end{equation}}

\begin{document}

\title{Neutrino Physics: Status and Prospects}

\author{K. SCHOLBERG}

\address{Massachusetts Institute of Technology, \\
Dept. of Physics, \\ 
Cambridge, MA 02139, USA\\ 
E-mail: schol@mit.edu}

\maketitle

\abstracts{This pedagogical overview will cover the current status
of neutrino physics from an experimentalist's point of view, focusing
primarily on oscillation studies. The evidence for neutrino
oscillations will be presented, along with the prospects for further
refinement of observations in each of the indicated regions of
two-flavor oscillation parameter space.  The next steps in oscillation
physics will then be covered (under the assumption of three-flavor
mixing): the quest for $\theta_{13}$, mass hierarchy and, eventually,
leptonic CP violation.  Prospects for non-oscillation aspects of
neutrino physics, such as kinematic tests for absolute neutrino mass
and double beta decay searches, will also be discussed briefly.}

\section{Neutrinos and Weak Interactions}

\begin{figure}[ht]
\begin{center}
\begin{tabular}{cccccccccccccccc}    
\begin{fmffile}{one}
  \fmfframe(1,7)(1,7){ 
   \begin{fmfgraph*}(110,80)
    \fmfleft{i1,i2}
    \fmfright{o1,o2}
    \fmflabel{$\nu_e$}{i1}
    \fmflabel{$u$}{i2}
    \fmflabel{$e^+$}{o1}
    \fmflabel{$d$}{o2}
    \fmf{fermion}{i1,v1,o1}
    \fmf{fermion}{i2,v2,o2}
    \fmf{scalar,label=$W^-$}{v1,v2}
   \end{fmfgraph*}
  }
\end{fmffile}

&&&&

\begin{fmffile}{two}
  \fmfframe(1,7)(1,7){ 
   \begin{fmfgraph*}(110,80)
    \fmfleft{i1,i2}
    \fmfright{o1,o2}
    \fmflabel{$\nu_x$}{i1}
    \fmflabel{$q$}{i2}
    \fmflabel{$\nu_x$}{o1}
    \fmflabel{$q$}{o2}
    \fmf{fermion}{i1,v1,o1}
    \fmf{fermion}{i2,v2,o2}
    \fmf{scalar,label=$Z^0$}{v1,v2}
   \end{fmfgraph*}
  }
\end{fmffile}

\end{tabular}

\caption{Examples of CC (left) and NC (right) neutrino interactions.\label{fig:ccnc}}
\end{center}
\end{figure}
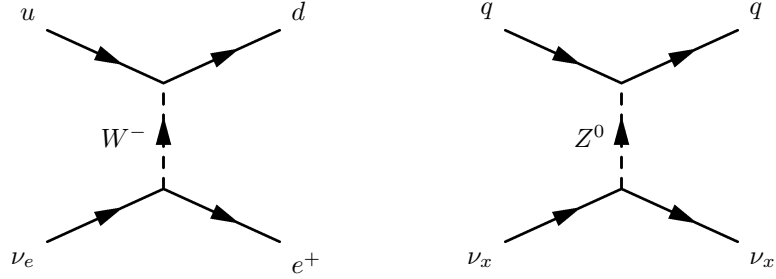

Neutrinos, the lightest of the fundamental fermions, are the neutral
partners to the charged leptons.  In the current picture, they come in
three flavors ($e$, $\mu$, $\tau$)\footnote{The $\tau$ neutrino has
only recently been directly detected by the DONUT
experiment\cite{donut}.} and interact only via the weak interaction.
In the Standard Model of particle physics, neutrinos interact with
matter in two ways: in a \textit{charged current} interaction, the
neutrino exchanges a charged $W^{\pm}$ boson with quarks (or leptons),
producing a lepton of the same flavor as the interacting neutrino
(assuming there is enough energy available to create the lepton.)  In
a \textit{neutral current} interaction, a neutral $Z$ boson is
exchanged; this type of interaction is
\textit{flavor-blind},
\textit{i.e.} the rate does not depend on the flavor of neutrino 
(see Figure~\ref{fig:ccnc}.)

\section{Neutrino Mass and Oscillations}\label{neutrinos}

Neutrinos are known to be very much lighter than their charged
lepton partners; direct measurements of neutrino mass
yield only upper limits of $<2$ eV$/c^2$.  However, the question
of neutrino mass can be probed using the oscillatory behavior 
of free-propagating neutrinos, which is dependent on the 
existence of non-zero neutrino mass.

Neutrino oscillations arise from straightforward quantum 
mechanics.  We assume that the $N$ neutrino flavor states 
$\ket{\nu_f}$, which participate
in the weak interactions, are superpositions of the mass
states $\ket{\nu_i}$, and are related by the
Maki-Nakagawa-Sakata (MNS) unitary mixing matrix:
\begin{equation}
\ket{\nu_f} = \sum_{i=1}^{N} U_{fi} \ket{\nu_i}.
\end{equation}\label{eq:nuosc}
For the two-flavor case, assuming
relativistic neutrinos, it can easily be shown that
the probability for flavor transition is given by
\begin{equation}\label{eq:oscprob}
P(\nu_f\rightarrow\nu_g)=1-|<\nu_f|\nu_g>|^2 = \sin^22\theta\sin^2(1.27\Delta m^2 L/E),
\end{equation}
for $\Delta m^2 \equiv m_2^2-m_1^2$ (in eV$^2$) and with $\theta$ the
angle of rotation.
$L$ (in km) is the distance traveled by the neutrino and $E$ (in GeV)
is its energy.  

Several comments are in order:

\begin{itemize}
\item Note that in this equation
the \textit{parameters of nature} that experimenters try to measure
(and theorists try to derive)
are $\sin^22\theta$ and $\Delta m^2$.
$L$ and $E$ depend on the experimental situation.  
\item The
neutrino oscillation probability depends on mass squared
differences, \textit{not} absolute masses.

\item In the three-flavor picture, the transition probabilities can be
computed in a straightforward way.  The
flavor states are related to the mass states according to

\begin{equation}
\left(
\begin{array}{c}
\nu_e  \\ \nu_{\mu} \\ \nu_{\tau} 
\end{array} \right)
= \left(
\begin{array}{ccc}
\rm{U}_{e1} & \rm{U}_{e2} & \rm{U}_{e3} \\ 
\rm{U}_{\mu 1} & \rm{U}_{\mu 2} & \rm{U}_{\mu 3} \\  
\rm{U}_{\tau 1} & \rm{U}_{\tau 2} & \rm{U}_{\tau 3} 
\end{array} \right)
\left(
\begin{array}{c}
\nu_1  \\ \nu_2 \\ \nu_3 
\end{array} \right)
\end{equation}  

and the transition probability is given by 
\begin{eqnarray*}
P(\nu_f\rightarrow\nu_g)  = \\ 
\delta_{fg} -4 \sum_{j>i}
& & {\rm Re}(\rm{U}^{\ast}_{fi} \rm{U}_{gi} \rm{U}_{fj}\rm{U}^{\ast}_{gj})\sin^2(1.27\Delta m_{ij}^2 L/E) \\
& & \pm 2 \sum_{j>i}\rm{Im}(\rm{U}_{fi}^{\ast} \rm{U}_{gi} \rm{U}_{fj}U^{\ast}_{gj})\sin^2(2.54\Delta m_{ij}^2 L/E),
\end{eqnarray*}
again for $L$ in km, $E$ in GeV, and $\Delta m^2$ in eV$^2$. The $-$
refers to neutrinos and the $+$ to antineutrinos.
\item For three mass states,
there are only two \textbf{independent} $\Delta m_{ij}^2$ values.  

\item If the mass states are not nearly degenerate,
one is often in a ``decoupled'' regime where it is possible to
describe the oscillation as effectively two-flavor, i.e. following
an equation similar to~\ref{eq:oscprob}, with effective mixing angles and
mass squared differences. We will assume a two-flavor description of the
mixing for most cases here.

\item ``Sterile'' neutrinos, $\nu_s$, with no normal
weak interactions, are possible in many theoretical scenarios (for
instance, as an isosinglet state in a GUT.)

\item When neutrinos propagate in matter, the oscillation probability
may be modified.  This modification is known as the the
``Mikheyev-Smirnov-Wolfenstein (MSW) effect'' or simply the ``matter
effect''.  Physically, neutrinos acquire effective masses via virtual
exchange of W bosons with matter (virtual CC interactions.)  For
example, consider $\nu_e$ propagating through solar matter: electron
neutrinos can
exchange W's with electrons in the medium, inducing an effective
potential $V=\sqrt{2} G_F N_e$, where $N_e$ is the electron density.
Muon and $\tau$-flavor neutrinos, however, can exchange virtual $Z$ bosons
only with the matter (because there are no $\mu$'s and $\tau$'s present.)  The
probability of flavor transition may be either enhanced or suppressed
in a way which depends on the density of matter traversed (and on the
vacuum oscillation parameters.)  
A description of the phenomenology of neutrino matter effects
may be found in \textit{e.g.} References~\cite{bahcallbook,bnv}.  We
will see below that matter effects become important for the solar neutrino
oscillation case, and also for future long baseline experiments.

\end{itemize}

\subsection{The Experimental Game}

The basic experiment to search for neutrino oscillations can
be described very simply.  
\begin{enumerate}

\item Start with some source
of neutrinos, either natural or artificial.

\item Calculate (or better yet, measure) the flavor
composition and energy spectrum of neutrinos.

\item Let the neutrinos propagate.

\item Measure the flavor composition and energy spectrum
after propagation.   Have the flavors and energies changed?  If
so, is the change described by the oscillation equation~\ref{eq:oscprob}?
And if so, what are the allowed parameters?

\end{enumerate}

The signature of neutrino oscillation manifests itself in one of
two ways, either by disappearance or appearance. In ``disappearance''
experiments, neutrinos appear to be lost as they propagate,
because they oscillate into some flavor with a lower interaction 
cross-section with matter.  An example of disappearance is a solar
neutrino experiment, for which $\nu_e$ transform
into muon/tau flavor neutrinos, which are below CC interaction
threshold at solar neutrino energies of a few MeV (solar $\nu_e$'s
do not have enough energy to create $\mu$ or $\tau$ leptons.)
In ``appearance'' experiments, one directly observes neutrinos of 
a flavor not present in the original source.  For example, one
might observe $\tau$'s from $\nu_{\tau}$ in a beam of multi-GeV $\nu_{\mu}$.


\section{The Experimental Evidence}
\noindent

There are currently three experimental indications of neutrino
oscillations. These indications are summarized in Table 1.  We will now
examine the current status of each of these observations.

\begin{table}[htbp]
\caption{Experimental evidence for neutrino oscillations.}
\centerline{\footnotesize
\begin{tabular}{c c c c c c}\\
\hline
$\nu$ source & Experiments & Flavors & $E$ & $L$  & $\Delta m^2$ sensitivity (eV$^2$)\\
\hline
Sun & Chlorine & $\nu_e \rightarrow \nu_x$ & 5-15 MeV & 10$^{8}$ km &  $10^{-12}-10^{-10}$\\
    & Gallium  &                           &          &           & or $10^{-6}-10^{-3}$ \\
    & Water Cherenkov &                    &          &           & \\
Reactor & Scintillator & $\bar{\nu}_e \rightarrow \bar{\nu}_x$ & 3-6 MeV & $\sim$180 km & $10^{-5}-10^{-3}$ \\
\hline
Cosmic ray & Water Cherenkov & $\nu_{\mu} \rightarrow \nu_x$ & 0.1-100 GeV & $10-10^5$ km & $10^{-2}-10^{-3}$ \\
showers   & Iron calorimeter & & & & \\
          & Upward muons & & & & \\
\hline
Accelerator & LSND & $\bar{\nu}_{\mu} \rightarrow \bar{\nu}_e$   & 15-50 MeV & 30 m & 0.1-1 \\ \hline
\end{tabular}}
\end{table}

\subsection{Atmospheric Neutrinos}
\noindent

Atmospheric neutrinos are produced by collisions of cosmic rays (which
are mostly protons) with the upper atmosphere.  Neutrino energies
range from about 0.1~GeV to 100~GeV.  At neutrino energies $\gsim$
1~GeV, for which the geomagnetic field has very little effect on the
primary cosmic rays, by geometry the neutrino flux should be up-down
symmetric.  Although the absolute flux prediction has $\sim$15\%
uncertainty, the flavor ratio (about two muon neutrinos for every
electron neutrino) is known quite robustly, since it depends on the
well-understood decay chain $\pi^{\pm} \rightarrow \mu^{\pm}~
\nu_\mu (\bar{\nu}_\mu) \rightarrow e^{\pm} \nu_e (\bar{\nu}_e)
\bar{\nu}_\mu (\nu_\mu)$.  The experimental strategy is to observe
high energy interactions of atmospheric neutrinos, tagging the flavor
of the incoming neutrino by the flavor of the outgoing lepton, which
can be determined from the pattern of energy loss: muons yield 
clean tracks, whereas high-energy electrons shower.  Furthermore, the
direction of the produced lepton follows the direction of the incoming
neutrino, so that the angular distribution reflects the neutrino
pathlength distribution.

\begin{figure}[ht]
\begin{minipage}[t]{2.2in}
\begin{centering}
\epsfxsize=2in \epsfbox{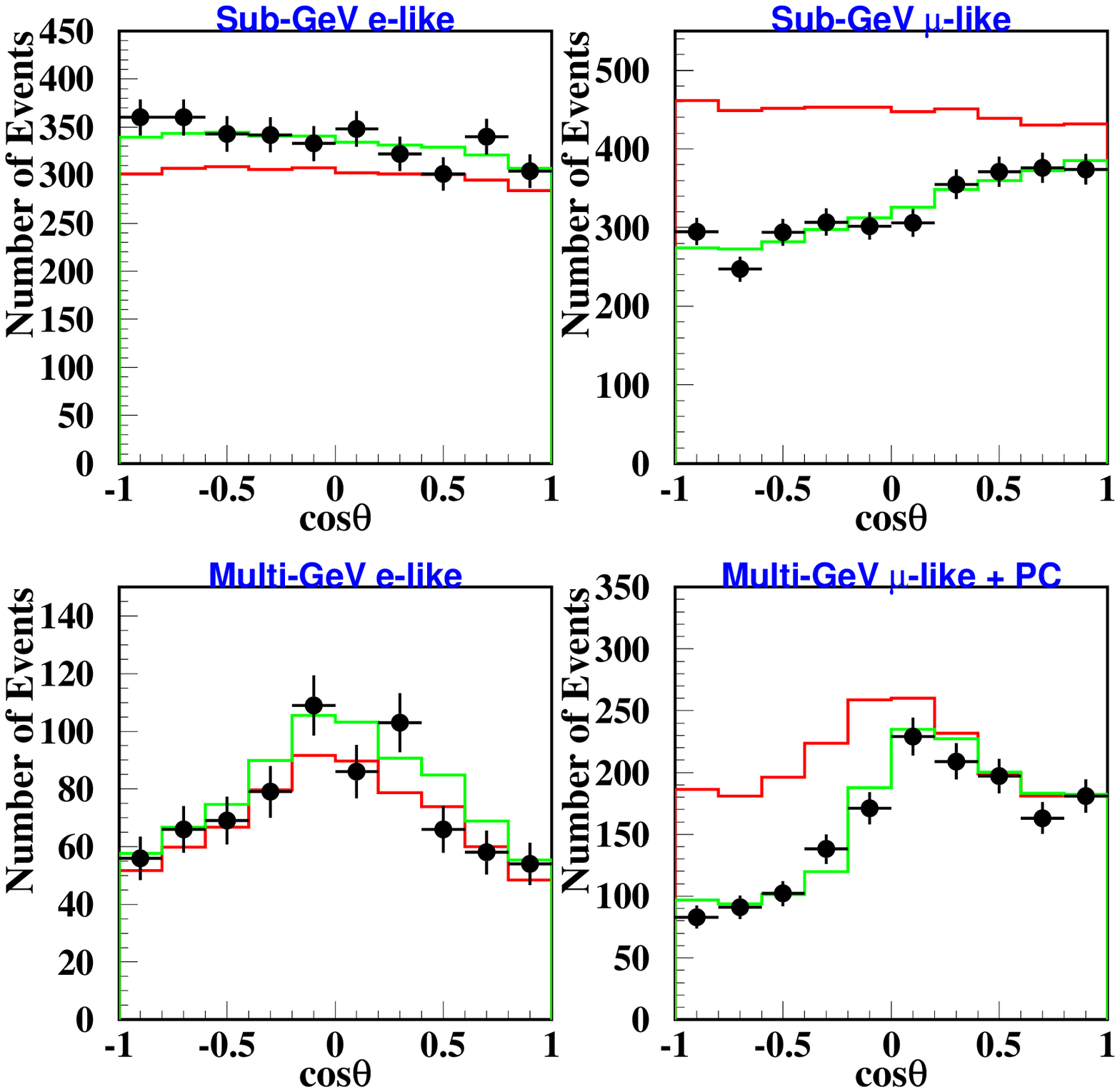}
\end{centering}
\end{minipage}
\begin{minipage}[t]{2.2in}
\begin{centering}
\epsfxsize=2in \epsfbox{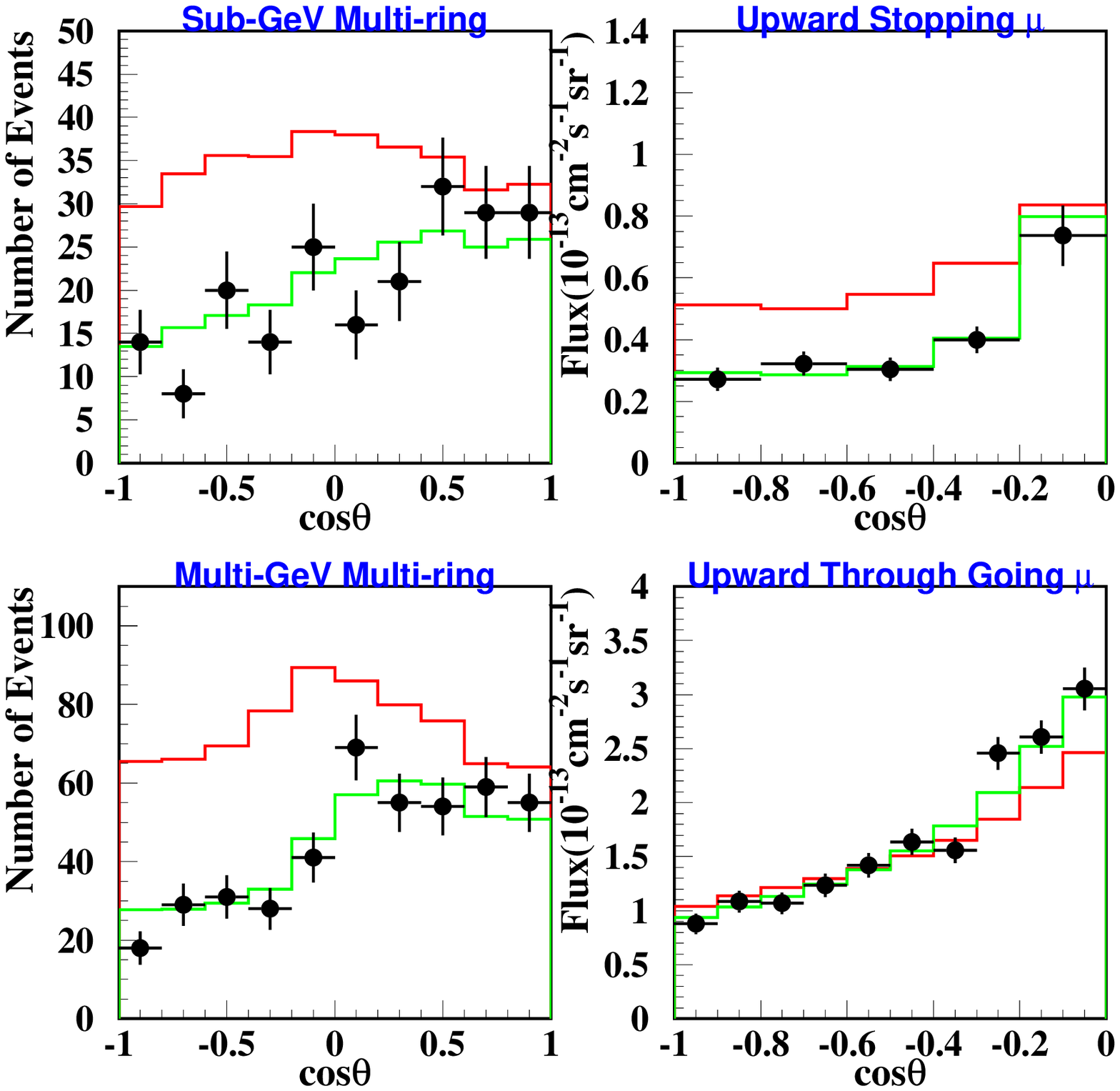}
\end{centering}
\end{minipage}
\caption{Zenith angle distributions for Super-K's 
newest 1489 day atmospheric neutrino
samples, including fully-contained events (those with interaction
products that do not leave the detector) and partially-contained events
(events with an exiting muon), upward through-going and stopping muons
(neutrinos interacting below the detector), and multiple ring events
(e.g. CC and NC single and multiple pion producing events.) The points
with (statistical) error bars are the data; the solid red line represents
the MC prediction for no oscillation; the paler green
line is the best fit for
$\nu_\mu \rightarrow \nu_{\tau}$ oscillation.  
\label{fig:skatmnu}}
\end{figure}

\begin{figure}[ht]
\centerline
{\epsfxsize=8cm   
\epsfbox{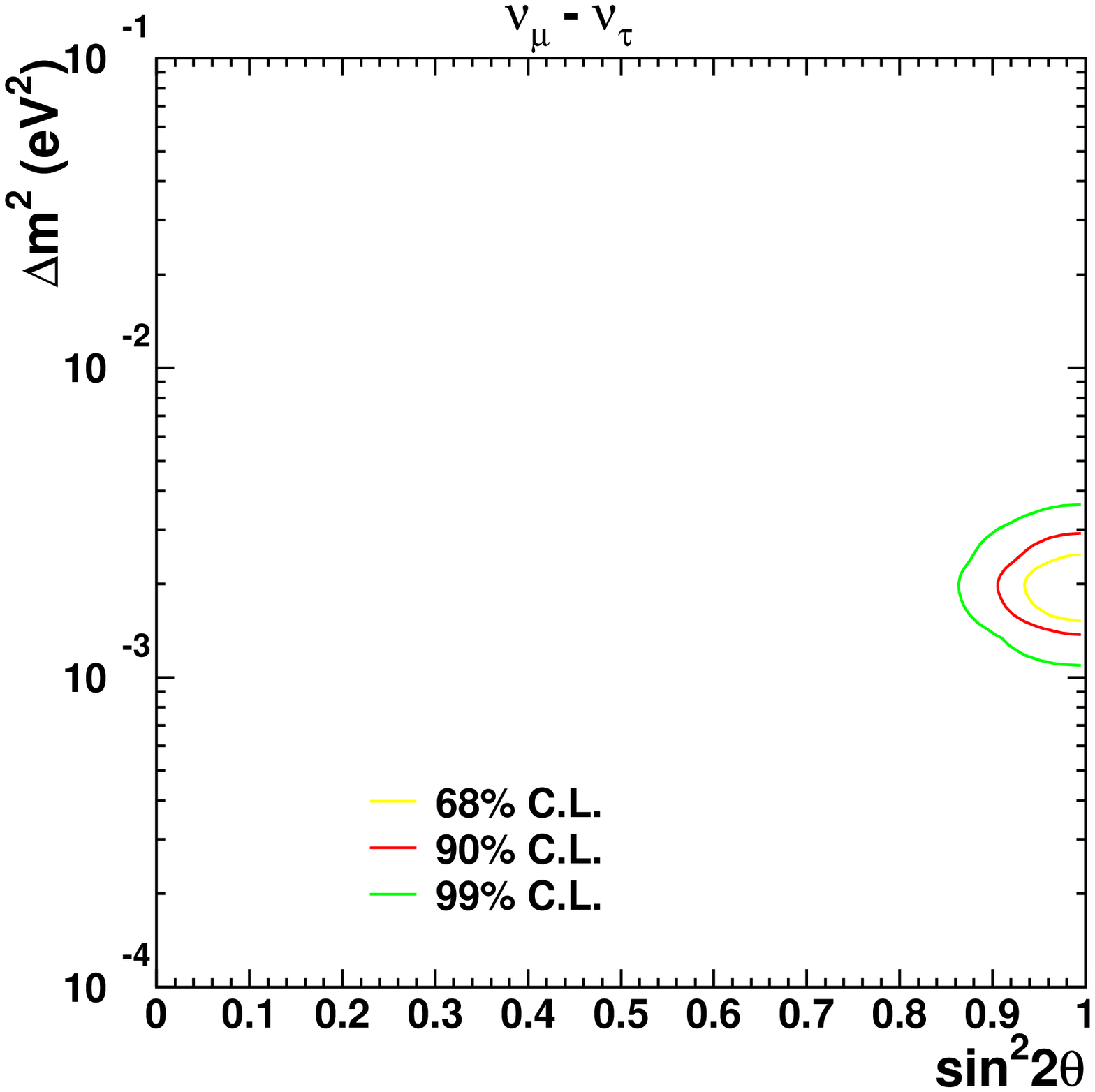}}
\caption{Right: allowed
region in oscillation parameter space corresponding to
the fit to Super-K atmospheric neutrino data (including
fully-contained events, partially-contained events, and upward-going
muons.) \label{fig:skoscparam}}
\end{figure}

Super-Kamiokande\cite{sktech}, a large water Cherenkov detector in
Japan, has shown a highly significant deficit of $\nu_\mu$ events from
below\cite{nuosc}, with an energy and pathlength dependence as expected from
equation~\ref{eq:oscprob} (see Figure~\ref{fig:skatmnu}.)  The most recent data constrain the
two-flavor $\nu_{\mu}\rightarrow \nu_{\tau}$ oscillation parameters to
a region as shown in Figure~\ref{fig:skoscparam}.  The latest results
from Soudan 2\cite{soudan} (an iron tracker) and from
MACRO's\cite{macro} upward-going muon sample are consistent with
the Super-K data.

Super-K has also been able to shed some light on the
flavors  involved in the atmospheric $\nu_{\mu}$ disappearance.  Assuming a
two-flavor oscillation, the missing $\nu_{\mu}$'s could have
oscillated into either $\nu_e$, $\nu_{\tau}$ or $\nu_s$.  The
oscillation cannot be pure $\nu_{\mu} \rightarrow \nu_{e}$, because
there is no significant excess of $\nu_e$ from below.  In addition,
the CHOOZ\cite{chooz} and Palo Verde\cite{paloverde} experiments
have ruled out disappearance of reactor $\bar{\nu}_e$; only 
small mixing to $\nu_e$ is allowed\cite{shiozawa}. (see Section~\ref{theta13}.)\footnote{In fact, a potential small $\nu_{\mu} \rightarrow \nu_{e}$ mixing
is extremely interesting,
as we will see in Section~\ref{beyond}.}

The $\nu_{\mu}\rightarrow\nu_{\tau}$ hypothesis is difficult to test
directly. Super-K expects relatively few charged current (CC)
$\nu_{\tau}$ interactions, and the products of
such interactions in the detector are nearly indistinguishable from
other atmospheric neutrino events.  However, recently Super-K has
employed several strategies to distinguish
$\nu_{\mu}\rightarrow\nu_{\tau}$ from
$\nu_{\mu}\rightarrow\nu_{s}$\cite{tausterile}.  First, one can look
for an angular distortion of high-energy neutrinos due to matter
effects of sterile neutrinos propagating in the Earth: unlike
$\nu_{\tau}$'s, sterile neutrinos do not exchange $Z^0$'s with matter
in the Earth, resulting in an matter effect that effectively
suppresses oscillation.  The effect is more pronounced at higher
energies.  Such distortion of the high-energy event angular distribution 
is not observed.  Second, one can
look at neutral current (NC) events in the detector: if oscillation is
to a sterile neutrino, the neutrinos ``really disappear'' and do not
interact via NC.  A NC-enriched sample of multiple-ring Super-K events
shows no deficit of up-going NC events.  Together, these measurements
exclude two-flavor $\nu_{\mu}\rightarrow\nu_{s}$ at 99\% C. L., for
all parameters allowed by the Super-K fully-contained events\cite{tausterile}.
The maximum allowed admixture of sterile neutrinos is about 20\%\cite{shiozawa}.

There is one more piece of evidence from Super-K suggesting that
$\nu_{\mu}\rightarrow\nu_{\tau}$ oscillations are primarily
responsible for the observed disappearance\cite{shiozawa,toshi}.
Because the energy threshold for tau production is about 3.5~GeV
and only a small fraction of the atmospheric neutrino flux exceeds this
energy,
only about 90 $\nu_{\tau}$-induced
$\tau$ leptons are expected in Super-K's 1489 day sample, given
the measured oscillation parameters.
Tau leptons decay with a very short lifetime into
a variety of modes, and can be observed as energetic multi-ring events;
such events are very difficult to disentangle from a large
background of multi-ring CC and NC events.
Nevertheless, three independent Super-K analyses 
which select ``$\tau$-like'' events have determined excesses of
up-going $\nu_{\tau}$ events consistent with $\tau$ appearance
at about the 2$\sigma$ level.

\subsubsection{Long Baseline Experiments}\label{longbaseline}

The next experiments to explore atmospheric neutrino parameter space are the
``long-baseline'' experiments, which aim to test the atmospheric
neutrino oscillation hypothesis directly with an artificial beam of
neutrinos.  In order to achieve sensitivity to the oscillation
parameters indicated by Super-K, $L/E$ must be such that for $\sim$1
GeV neutrinos, baselines are hundreds of kilometers.  A beam is
created by accelerating protons and bombarding a target to produce
pions and other hadrons; pions are then focused forward with a
high-current magnetic ``horn'' and allowed to decay in a long pipe.
The neutrino flavor composition and spectrum can be measured in a
near detector before propagation to a distant far detector.

\begin{figure}[ht]
\begin{minipage}[t]{2.2in}
\begin{centering}
\epsfxsize=2in \epsfbox{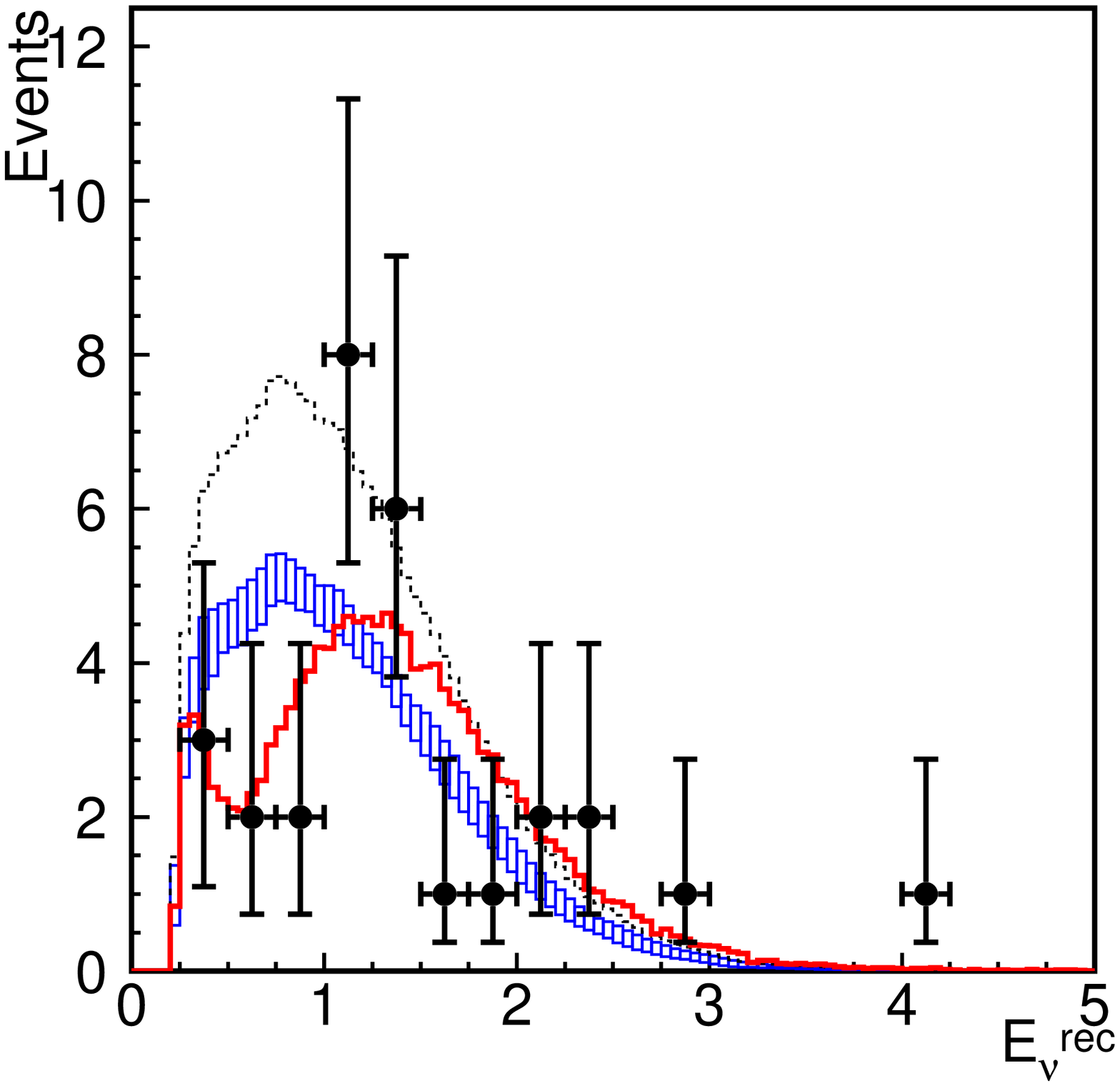}
\end{centering}
\end{minipage}
\begin{minipage}[t]{2.2in}
\begin{centering}
\epsfxsize=2in \epsfbox{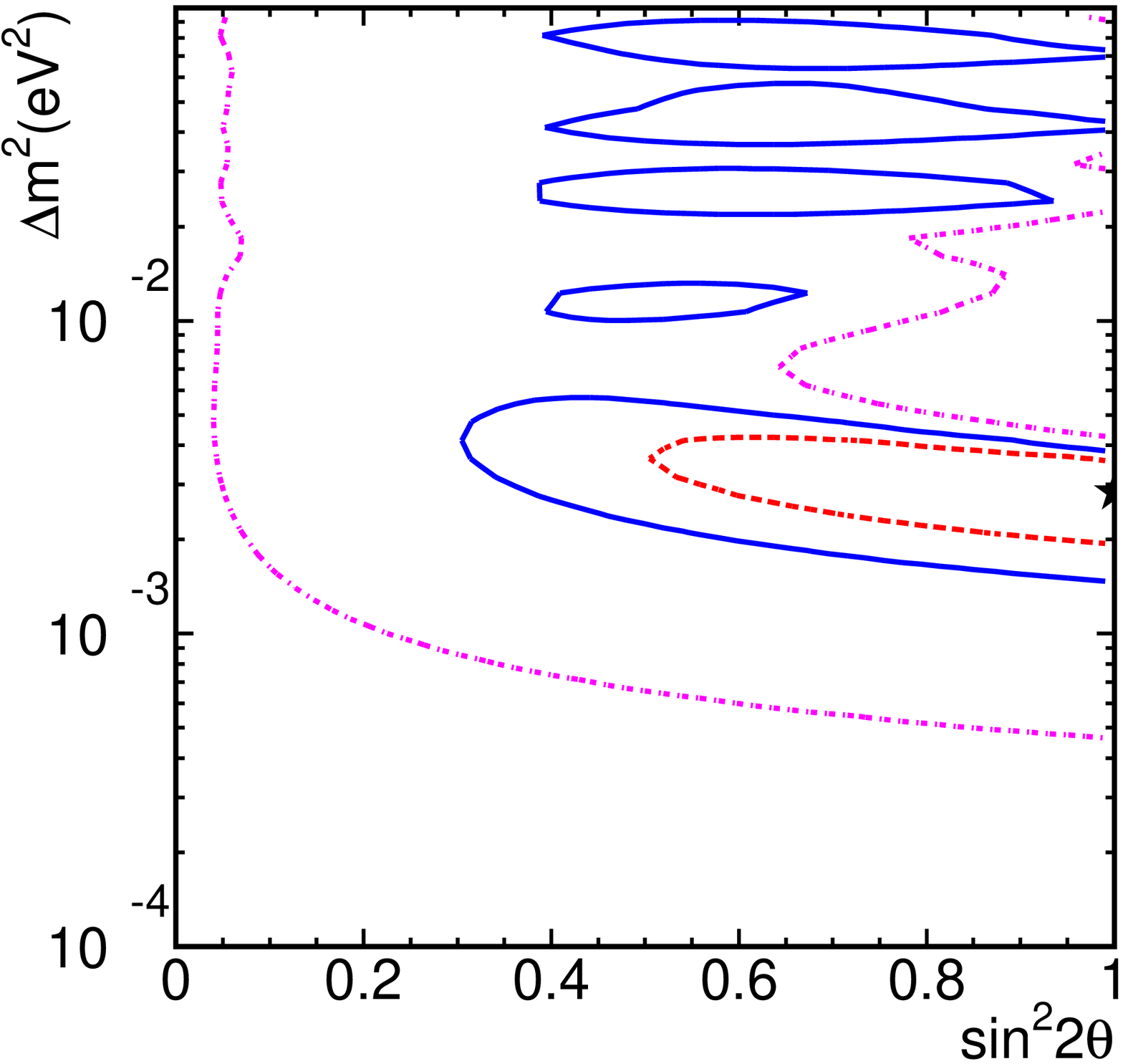}
\end{centering}
\end{minipage}
\caption{Left: Expected beam neutrino spectrum for no
oscillations (dashed
line), data (points), expected spectrum with systematic
error normalized to the number of observed events (boxes)
and best fit to the oscillation hypothesis
(solid) for the K2K 1999-2001 data sample. 
Right: allowed
region in oscillation parameter space corresponding
to K2K 1999-2001 data sample, using both
suppression and spectrum. \label{fig:k2kresult}}
\end{figure}

The first long-baseline experiment is the K2K (KEK to Kamioka)
experiment\cite{k2k}, which started in March 1999, and which saw the
first artificial long-distance neutrinos in June 1999.  K2K sends a
beam of $\langle{E}_{\nu}\rangle\sim$1~GeV $\nu_{\mu}$ 250~km across
Japan to the Super-K experiment. K2K can look for $\nu_{\mu}$
disappearance (the beam energy is not high enough to make significant
numbers of $\tau$'s.)  Preliminary K2K results\cite{k2k,k2kosc} do
show a deficit of observed neutrinos: 80.1$^{+6.2}_{-5.4}$ beam events
in the fiducial volume are expected, based on beam-modeling and near
detector measurements; however only 56 single-ring $\nu_{\mu}$ events
were seen at Super-K.  The far spectrum was also measured.  The best
fit oscillation parameters using both spectrum and suppression
information are entirely consistent with the atmospheric results.
See Figure~\ref{fig:k2kresult}.  Somewhat more than half of K2K data
has now been taken.  The beam resumed in early 2003 after repair of
Super-K.  The next generation long baseline experiments will be
discussed in Section~\ref{nextlb}.

\subsection{Solar Neutrinos}
\noindent

The deficit of solar neutrinos was the
first experimental hint of neutrino oscillations.  The solar neutrino energy
spectrum is well-predicted, and depends primarily on weak physics,
being rather insensitive to solar physics.  The three ``classic''
solar neutrino detectors (chlorine, gallium and water Cherenkov),
with sensitivity at three different
energy thresholds, together observe an energy-dependent suppression
which cannot be explained by any solar model 
(standard or non-standard)\cite{bahcall}.

\begin{figure}[ht]
\centerline
{\epsfxsize=4in\epsfbox{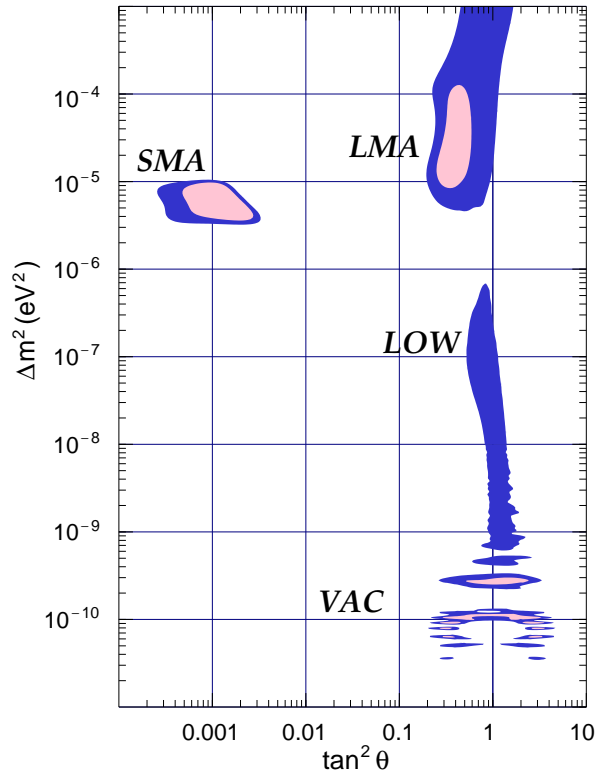}}
\caption{Solar neutrino parameter space: 
the shaded 
areas show the ``classic'' global flux fit solutions
from chlorine, gallium and water Cherenkov experiments
(from Reference~\protect\cite{murayama}.)
\label{fig:solar_param}}
\end{figure}

The observed suppression in all three experiments can be explained by
neutrino oscillation at certain values of $\Delta m^2$ and mixing
angle: see Figure~\ref{fig:solar_param}.  The ``classic'' allowed regions at
higher values of $\Delta m^2$ (``small mixing angle'', ``large mixing
angle'' and ``low'') are those for which matter effects in the Sun
come into play. There are also solutions at
very small $\Delta m^2$ values for which matter effects
in the sun are not involved: these are known as ``vacuum''
oscillation or ``just-so'' solutions.\footnote{The vacuum
solutions are ``just-so''
because 
oscillation parameters must fine-tuned to
explain suppression at exactly the Earth-Sun distance; on the other hand,
because the Sun has a range of electron densities, 
$\nu_e$ suppression will result for a broader range of 
oscillation parameters if one assumes that matter effects are involved.}
Figure~\ref{fig:solar_param} shows the mixing
angle axis plotted as $\tan^2\theta$, rather than as the more
conventional $\sin^22\theta$, to make evident the difference between
$0<\theta<\pi/4$ and $\pi/4<\theta<\pi/2$: these regions are not
equivalent when one considers matter effects\cite{murayama}.

\begin{figure}[ht]
\centerline
{\epsfxsize=4in\epsfbox{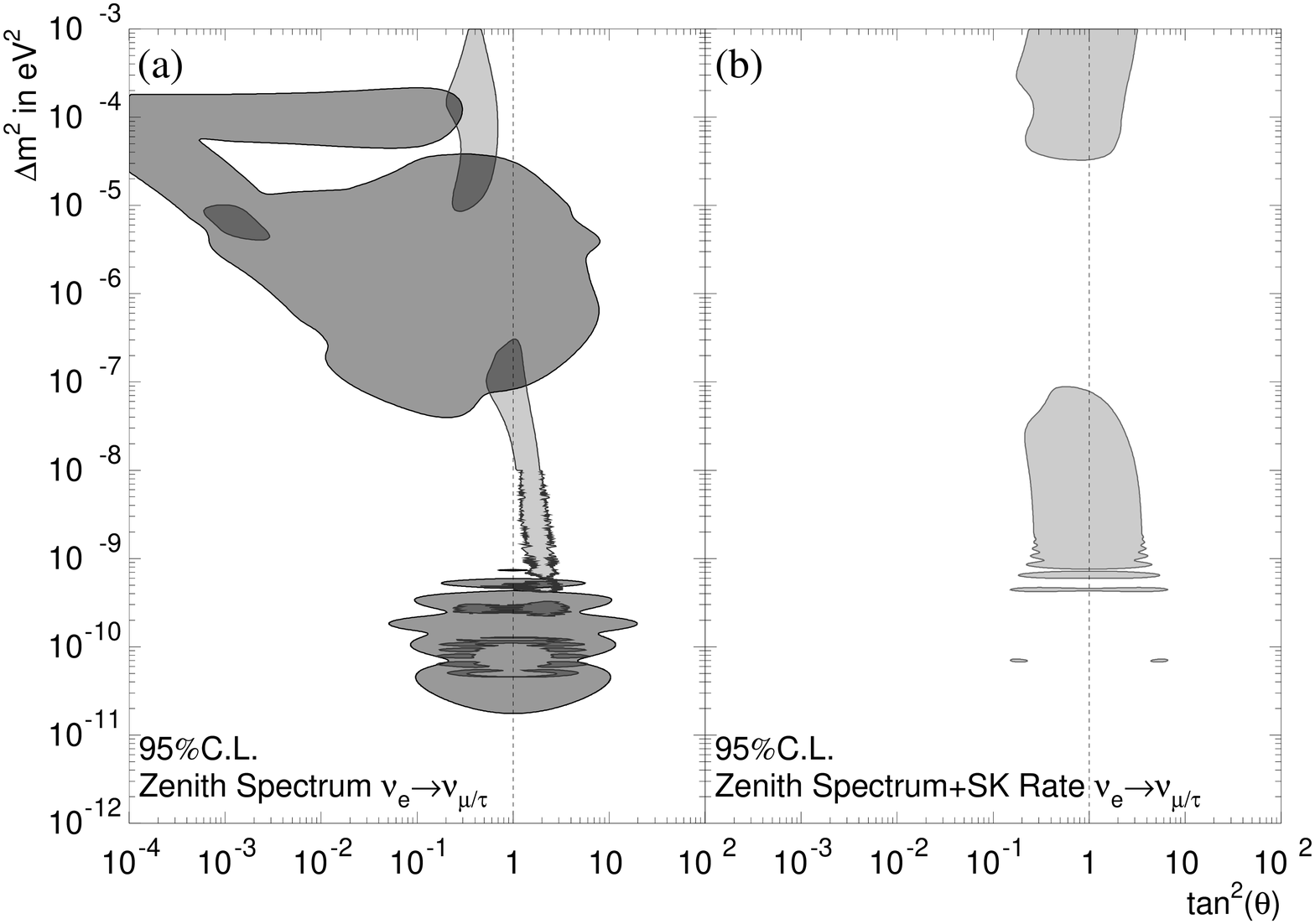}}
\caption{Solar neutrino parameter space: on the left,
the light grey 
areas show the ``classic'' global flux fit solutions
from chlorine, gallium and the SNO experiment's CC measurement.  
The 
darker grey shaded regions indicate Super-K's excluded
regions from spectral and day/night information (and the
darkest grey regions indicate the overlap.)
On the right, the light shaded areas indicate 
allowed regions from Super-K data alone and SSM $^8$B neutrino flux.\label{fig:sksolar_param}}
\end{figure}

Before 2000, the most precise real-time solar neutrino data came from
Super-K via the elastic scattering reaction $\nu_{e,x} + e^-
\rightarrow \nu_{e,x} + e^{-}$, which proceeds via both CC and NC
channels, with a cross-section ratio of about 1:6. In this reaction, the
Cherenkov light of the scattered electron is measured.  The scattered
electrons point away from the direction of the sun.
 
Possible
``smoking guns'' for neutrino oscillations include a distortion from the
expected shape that would be hard to explain by other than non-standard weak
physics.  The latest Super-K solar neutrino spectrum shows no evidence
for distortion\cite{sksolar}.  Another ``smoking gun'' solar neutrino
measurement is the day/night asymmetry: electron neutrinos may be
regenerated in the Earth from their oscillated state for certain
oscillation parameters.  The latest measured Super-K day/night
asymmetry is $\frac{day-night}{(day+night)/2}=-0.021 \pm 0.020 {\rm stat}
^{+0.013}_{-0.012}$ (syst): regeneration is therefore a relatively small
effect, if it is present at all.  Together, the energy
spectrum and day/night observations place
strong constraints on solar neutrino parameters.  In particular,
Figure~\ref{fig:solar_param} shows the Super-K results overlaid on the
global flux fit parameters: large mixing angles are favored, and the
small mixing angle and vacuum solutions from the global flux fit are
disfavored at 95\% C.L..  Global flux fit $\nu_e
\rightarrow \nu_s$ solutions are also disfavored.  

The information from Super-K served primarily to constrain parameters.
The true ``smoking gun'' for solar neutrino oscillations recently came
from the Sudbury Neutrino Observatory\cite{sno}, a detector
comprising 1~kton of D$_2$O in Sudbury, Canada, with the unique
capability to detect neutral current reactions from the breakup of
deuterium, $\nu_x + d \rightarrow \nu_x + p + n$: since this reaction
is flavor-blind, it measures the \textit{total} active neutrino flux
from the sun.  Neutrons from this reaction can be detected via various
methods: capture on $d$ itself, capture on Cl ions from dissolved
salt, and neutron detectors.  In addition, the charged current
reaction $\nu_e + d \rightarrow \nu_e + p + e^-$ specifically tags the
$\nu_e$ component of the solar flux.  SNO also observes the same
neutrino-electron elastic scattering (ES) interaction as Super-K,
which proceeds via both CC and NC channels.

\begin{figure}[ht]
\begin{minipage}[t]{2.2in}
\begin{centering}
\epsfxsize=2.2in \epsfbox{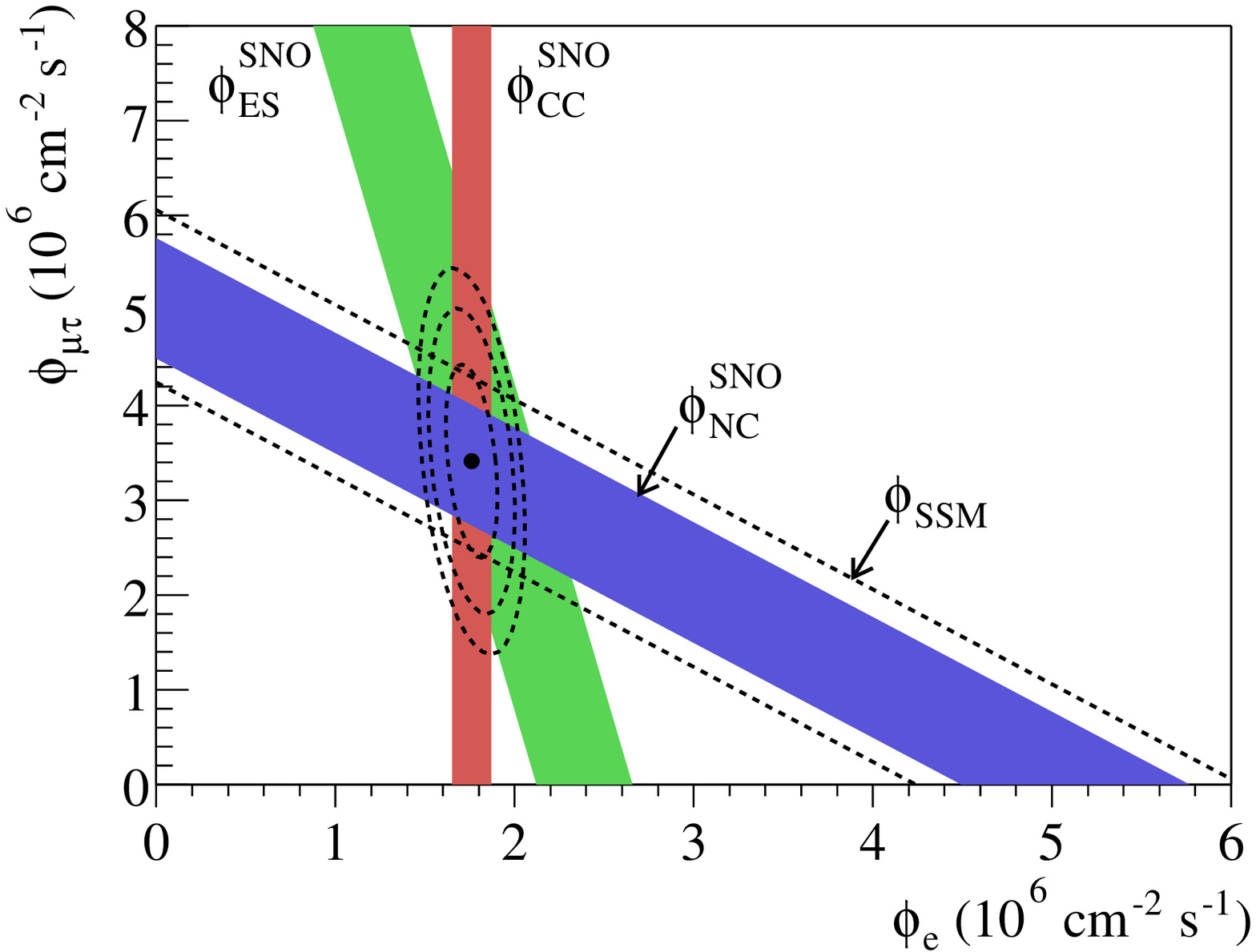}
\end{centering}
\end{minipage}
\begin{minipage}[t]{2.0in}
\begin{centering}
\epsfxsize=2.0in \epsfbox{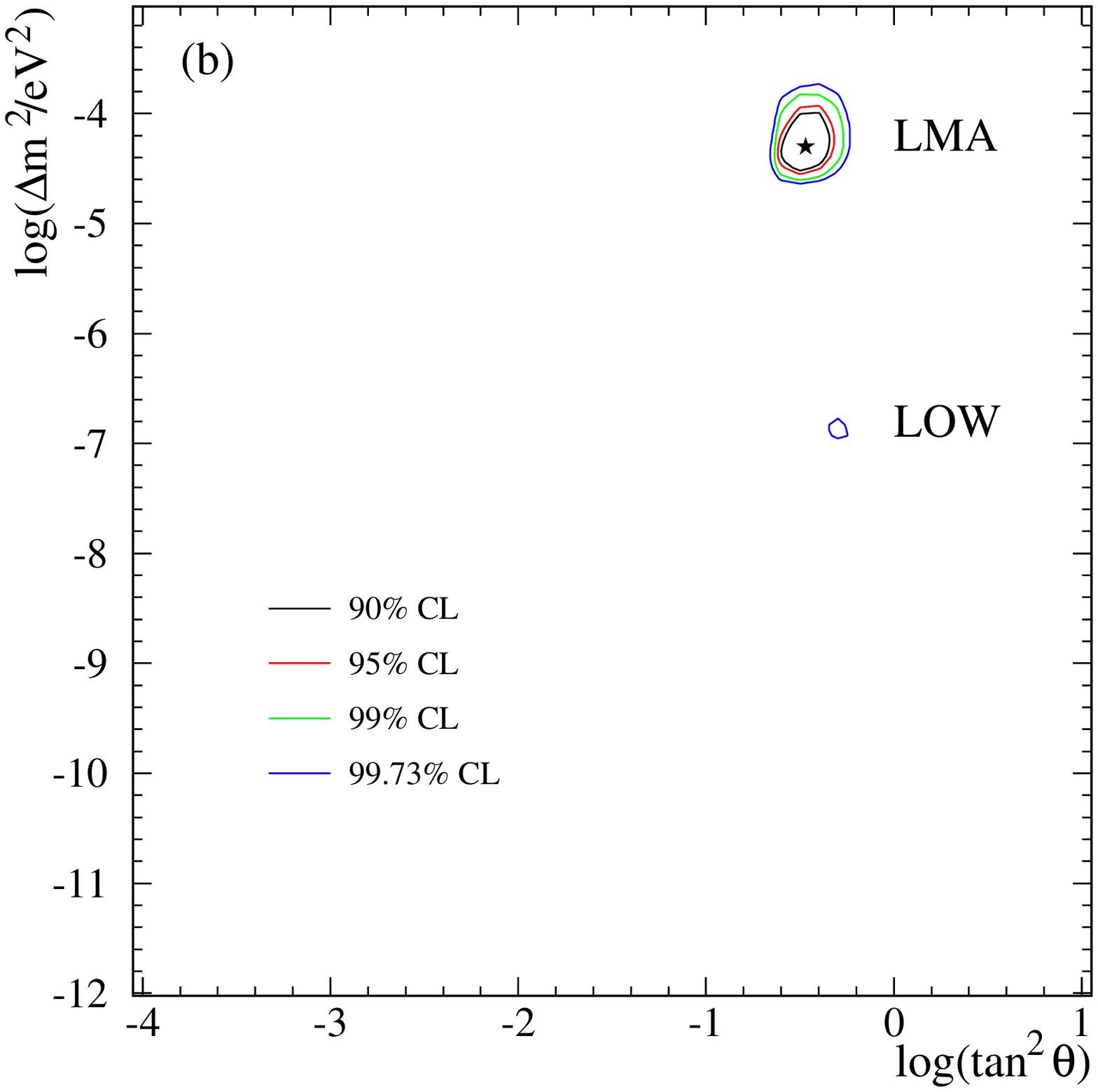}
\end{centering}
\end{minipage}
\caption{Left: Inferred flavor components from fluxes measured at SNO
(see text.) 
Right: allowed
region in oscillation parameter space after SNO 2002 results. \label{fig:snoresult}}
\end{figure}

SNO's recent results\cite{snodata} are summarized in
Figure~\ref{fig:snoresult}, which shows 
the measured fluxes $\phi_{\mu \tau}$ vs $\phi_{e}$.\footnote{Note
that one cannot distinguish between $\nu_{\mu}$ and $\nu_{\tau}$ at
low energy since the NC interaction does not distinguish between
them.} 
The CC measurement, which tags $\nu_e$ flux $\phi_e$, is
represented by a vertical bar on this plot.  Since the neutral current
flux is flavor-blind and therefore represents a measurement of the sum
of $\phi_{\mu \tau}$ and $\phi_{e}$, \textit{i.e.} $\phi_{NC} = \phi_{\mu
\tau}+\phi_{e}$, the NC measurement corresponds to a straight line
with slope $-1$ on this plot.  The intersection with the vertical CC
line indicates the composition of the solar neutrino flux: it is
approximately $1/3$ $\nu_e$ and $2/3$ $\nu_{\mu,\tau}$.
 The ES reaction measures both $\nu_e$ and $\nu_{\mu,\tau}$
with a known ratio, $\phi_{ES} = 0.154 \phi_{\mu \tau}+ \phi_{e}$ for SNO, so
that the ES measurement corresponds to a line on the plot with slope
$-1/0.154$; it provides a consistency check.  The conclusion from SNO is
that solar neutrinos \textit{really are oscillating} (into active
neutrinos.)  The solar neutrino problem is solved!

The detailed measurements from SNO incorporating observed day/night asymmetry
and energy spectra shrink the allowed parameters down to small regions,
shown in Figure~\ref{fig:snoresult}.  At 99\% C.L., only the LMA region
is left.  

So far, SNO's NC measurement comes from capture of neutrons
on $d$;  SNO continues to run, and will provide cross-checked NC
measurements using salt and helium neutron counters.

\begin{figure}[ht]
\begin{minipage}[t]{2.3in}
\begin{centering}
\epsfxsize=2.3in \epsfbox{reactor_v26.eps}
\end{centering}
\end{minipage}
\begin{minipage}[t]{2.0in}
\begin{centering}
\epsfxsize=2.0in \epsfbox{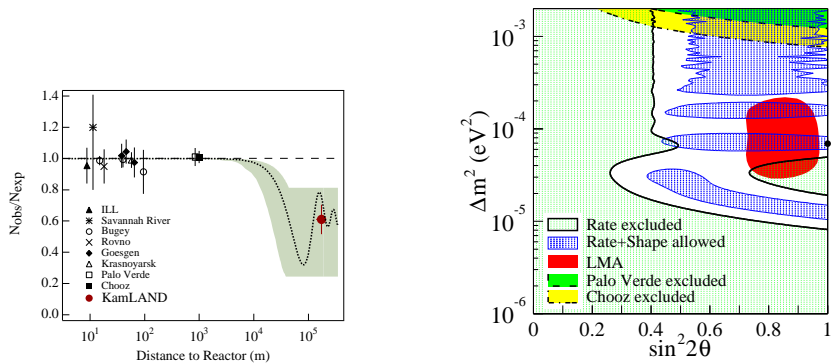}
\end{centering}
\end{minipage}
\caption{Left: Ratio of measured to expected $\bar{\nu}_e$ for various
experiments as a function of baseline; the point on the far right
is the KamLAND result.  The shaded area represents the expectation
from the solar LMA solution, and the dotted line is the best
fit to the oscillation hypothesis. 
Right:  Allowed and excluded regions in oscillation parameter
space for various experiments (as indicated in the legend.) 
\label{fig:kamlandresult}}
\end{figure}

There is one more recent chapter in the solar neutrino story.
KamLAND, a 1~kton scintillator detector at the Kamioka mine in
Japan\cite{kamland}, has investigated solar neutrino oscillation
parameters using reactor neutrinos rather than solar neutrinos
directly.  Reactors produce $\bar{\nu}_e$ of few-MeV energies
abundantly; assuming vacuum oscillations, the baseline required to observe
oscillations with LMA parameters is about $\sim$ 100~km.  Note that no
significant matter effects are expected at this baseline.  KamLAND
observes the sum of the fluxes of neutrinos from reactors in Japan and
Korea, with roughly a $180$~km average baseline, via the inverse beta
decay reaction $\bar{\nu}_e + p \rightarrow e^{+} + n$; 
$\bar{\nu}_e$'s are tagged using the coincidence
between the positron and the 2.2~MeV $\gamma$-ray from
the captured neutron.  In December
2002, the KamLAND experiment announced an observed suppression of
reaction $\bar{\nu}_e$ consistent with LMA parameters: 
see Figure~\ref{fig:kamlandresult}.  Solar neutrino
oscillations are therefore now confirmed using a completely
independent source of neutrinos and experimental technique.  In
addition, the LMA solution is strongly indicated.

\subsection{LSND}

\begin{figure}[ht]
\centerline{\epsfxsize=10cm   
\epsfbox{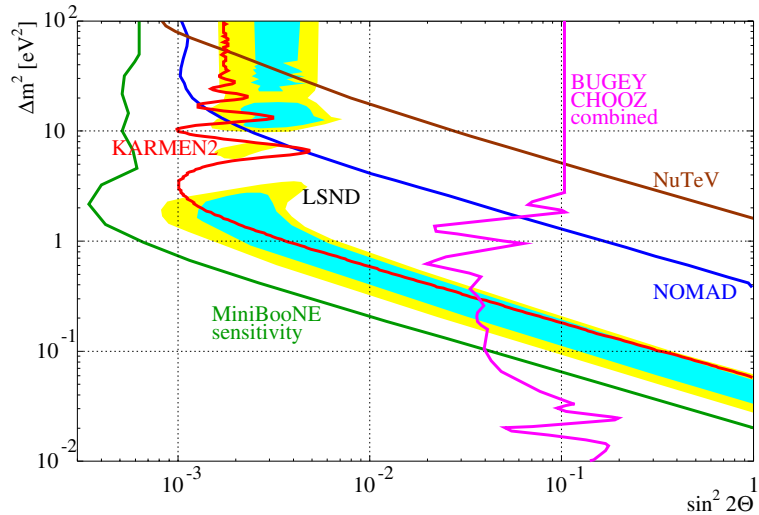}}
\caption{The shaded region shows the LSND allowed regions at 90\% and 99\% C.L.; the region to the right of the KARMEN2 line is excluded by KARMEN.  Also
shown are exclusions by the reactor experiments Bugey and Chooz, the NuTeV
and Nomad excluded regions, and the reach of the mini-BooNE experiment (see Section~\ref{lsndnext}.)\label{fig:lsnd_space}}
\end{figure}

The third oscillation hint is the only ``appearance'' observation: the
Liquid Scintillator Neutrino Detector (LSND) experiment at Los Alamos
has observed an excess of $\bar{\nu}_e$ events\cite{lsnd} from a
beam which should contain only $\bar{\nu}_{\mu}$, $\nu_e$ and
$\nu_{\mu}$ from positive pion and muon decay at rest.  The result is
interpreted as $\sim$20-50~MeV $\bar{\nu}_{\mu}$'s oscillating over a
30~m baseline.  See
Figure~\ref{fig:lsnd_space} for the corresponding allowed region in
parameter space, which is
at large $\Delta m^2$ and small mixing.
(The large mixing angle part of this range is ruled out
by reactor experiments.)

An experiment at Rutherford-Appleton Laboratories in the U. K. 
called KARMEN, which
has roughly similar neutrino oscillation sensitivity as does LSND
(although with a shorter 17.5~m baseline), does not however confirm
the LSND result\cite{karmen}.  This detector expects fewer signal
events than does LSND, but has a stronger background rejection due to
the pulsed nature of the ISIS neutrino source.  However, due to somewhat
different sensitivity, KARMEN's lack of observation of $\bar{\nu}_e$
appearance cannot rule out all of the parameter space indicated by
LSND: see Figure~\ref{fig:lsnd_space}.

\section{Where Do We Stand?}

Now we can step back and view the big picture.  Where do we stand?
The current experimental picture for the three oscillation signal
indications can be summarized:

\begin{itemize}

\item For atmospheric neutrino parameter space: evidence
from  Super-K, Soudan 2 and MACRO 
is very strong for $\nu_{\mu} \rightarrow \nu_x$.  Furthermore,
Super-K's data favor the $\nu_{\mu} \rightarrow \nu_{\tau}$
hypothesis over the $\nu_{\mu} \rightarrow \nu_s$ one.  These
oscillation parameters have been independently confirmed
using the K2K beam of $\sim$1~GeV $\nu_{\mu}$'s to Super-K.

\item For solar neutrino parameter space ($\nu_e \rightarrow \nu_x$):
The solar neutrino problem is now solved.  While Super-K data favored
large mixing via day/night and spectral measurements, SNO's
D$_2$O-based NC and CC measurements have confirmed that solar neutrinos
are oscillating, and have shrunk down the allowed parameter space to
the LMA region using day/night and spectral measurements.  Better yet,
the KamLAND experiment has independently confirmed the LMA solution
using reactor $\bar{\nu}_e$'s.  Oscillation to sterile neutrinos
is disfavored.

\item The LSND indication of $\bar{\nu}_{\mu} \rightarrow \bar{\nu}_e$ still
stands; KARMEN does not rule out all of LSND's allowed parameters.
\end{itemize}

\begin{figure}[ht]
\epsfxsize=10cm   
\epsfbox{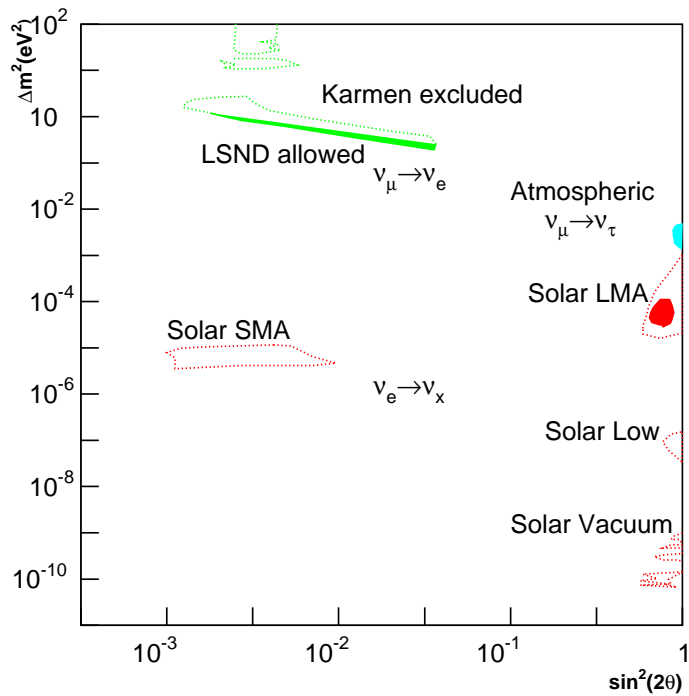}
\caption{Oscillation parameter space showing all three indications of
oscillation, in the two-flavor mixing approximation.  At high $\Delta
m^2$, the parameters allowed by LSND are shown by dotted lines, and
the part not excluded by Karmen is shown as a solid region.  Allowed
atmospheric neutrino parameters are shown at large mixing and $\Delta
m^2$ of about $2.5 \times 10^{-3}$ eV$^2$.  Also shown by dotted lines are
the ``classic'' solar neutrino solutions at small $\Delta m^2$:
``small mixing angle'' (SMA), ``large mixing angle''(LMA), and
``low'', which all involve matter effects in the sun, and the vacuum
solutions at very small $\Delta m^2$.  With new information from SNO
and KamLAND, only the LMA solution is now allowed, as indicated by the
solid region at about $4.5\times 10^{-4}$ eV$^2$.\label{fig:big_picture}}
\end{figure}

What do these data mean?  There is an obvious problem. Under the
assumption of three generations of massive neutrinos, there are only
two independent values of $\Delta m_{ij}^2$: we must have $\Delta
m_{13}^2=\Delta m_{12}^2 + \Delta m_{23}^2$.  However, we have three
measurements which give $\Delta m_{ij}^2$ values of three different
orders of magnitude.  So, if each hint represents two-flavor mixing,
then something must be wrong.  All data cannot be satisfactorily
fit assuming three-flavor oscillations.
One way to wriggle out of this difficulty is to introduce another
degree of freedom in the form of a sterile neutrino (or neutrinos) or
else invoke some exotic solution (\textit{e.g.} CPT
violation\cite{cpt}.)  (We cannot introduce another active
neutrino, due to the Z$^0$ width measurements from LEP, which
constrain the number of light active neutrinos to be three\cite{lep}: any
new light neutrino must be sterile.)  Although pure mixing into
$\nu_s$ is now disfavored by solar and atmospheric neutrino results, a
sterile neutrino is still barely viable as part of some four-flavor
mixing\cite{4flavfit}.  Of course, it is also possible that some of
the data are wrong or misinterpreted.  Clearly, we need more
experiments to clarify the situation.

\section{What's Next for Two-flavor Oscillations?}

So what's next?  First, let's consider the next experiments
for each of the interesting regions of two-flavor parameter space.

\subsection{LSND Neutrino Parameter Space}\label{lsndnext}

The next experiment to investigate the LSND parameter space will be
BooNE (Booster Neutrino Experiment.) This will look at $\sim$ 1 GeV
neutrinos from the 8~GeV booster at Fermilab, at a baseline of about
500~m (with a second experiment planned at longer baseline if an
oscillation signal is seen.)  This experiment 
is primarily designed to test $\nu_{\mu}\rightarrow
\nu_e$ at about the same $L/E$ as LSND. Since the neutrino energy is higher,
and the backgrounds are different, systematics will presumably be
different from those at LSND.  BooNE, which started in 2002, expects
to cover all of LSND parameter space\cite{boone} (see
Figure~\ref{fig:lsnd_space}.)  If a signal is found, the BooNE
collaboration plans to build another detector at a longer baseline to
further test the oscillation hypothesis.

\subsection{Solar Neutrino Parameter Space}

Now that the latest results from SNO and KamLAND have squeezed the
allowed solar mixing parameters down to the LMA region, solar neutrino
physics is entering a precision measurement era.  Over the next few
years, we expect to have cross-checks of NC measurements from SNO,
using different neutron detection techniques (salt, NCDs.)  From
KamLAND we expect better precision from improved statistics and
systematics; KamLAND will also attempt to measure the solar neutrino
flux directly.  Borexino, a planned 300~ton scintillator experiment at
the Gran Sasso Laboratory in Italy\cite{borexino} with very low
radioactive background, hopes to measure the solar $^7$Be line
at 0.86~MeV.

The true frontier for solar neutrino experiments is the real-time,
spectral measurement of the flux of neutrinos below 0.4~MeV produced
by pp reactions in the sun, which are responsible for most of the
solar energy generation.  The pp flux is precisely known, which will
aid in precision measurements of mixing parameters; in addition if the
total pp flux is well-known, measurement of the active component will
help constrain a possible sterile admixture. The pp flux is also a new
window on solar energy generation.  Because the pp flux is very large,
one can build relatively small (tens to hundreds of tons) detectors
and still expect a reasonable rate of neutrino interactions.  The
challenge is to achieve low background at low energy threshold.  There
are a number of innovative new solar neutrino experiments aiming to
look at the very low energy pp solar flux\cite{lownu}, among them
LENS, Heron, solar-TPC and CLEAN.

\subsection{Atmospheric Neutrino Parameter Space}\label{nextlb}

Two-flavor oscillation studies at atmospheric neutrino parameters has also
entered a precision measurement era.

The K2K experiment will continue, now that Super-K has been
refurbished to 47\% of its original number of inner detector phototubes
after the accident of November 2001.  The results published so
far represent about half of the total number of protons on target
for the neutrino beam; the next few years will see both systematic
and statistical precision improvements in mixing parameter measurements.

The next set of long baseline experiments to explore atmospheric
oscillation parameter space have $\sim$730~km baselines
and will start in a few years.  The NuMi 
beamline\cite{numi} will send a $\nu_{\mu}$
beam from Fermilab to Soudan, with a beam energy of $3-8$ GeV, and a
baseline of 735~km.  The far detector, MINOS\cite{minos} is a magnetic iron tracker. A primary 
goal is to attain 10\% precision on 2-3 mixing parameters
$\Delta m^2_{23}$ and $\sin^2 2\theta_{23}$.

CNGS (Cern Neutrinos to Gran Sasso)\cite{cngs} is a $\sim$20~GeV
$\nu_{\mu}$ beam from CERN to the Gran Sasso 730~km away.  
The two planned CNGS detectors, OPERA\cite{opera} and
Icarus\cite{icarus}, are focused on an explicit $\nu_{\tau}$
appearance search. Because when $\tau$'s decay they make tracks only
about 1~mm long, both detectors are fine-grained imagers.  Icarus is a
liquid argon time projection chamber, and OPERA is a hybrid
emulsion/scintillator detector.  Both experiments expect a few dozen
$\tau$ events over several years of running.

\section{Beyond Two-Flavor Oscillations}\label{beyond}

The previous section discussed the future of neutrino oscillation
studies in the context of two-flavor oscillations.  As noted
in section~\ref{neutrinos}, however, this is an approximation valid
for well-separated mass states, which appears to be the case.
However a full description requires three flavors.

In the following, we will assume that a ``standard'' three-flavor picture is valid.
If mini-BooNE confirms the LSND effect, we will have to rethink
our picture, and our goals.

In this ``standard'' picture, neutrino mixing can be described by
six parameters:  two independent $\Delta m_{ij}^2$
($\Delta m^2_{12}$, $\Delta m^2_{23}$), three mixing angles
($\theta_{12}$, $\theta_{23}$, $\theta_{13}$), and
a CP violating phase $\delta$.\footnote{Majorana phases, which
cannot be measured in oscillation experiments and in general
are very difficult to observe~\cite{majorana}, will not be considered here.}
The mixing matrix U of equation~\ref{eq:nuosc} can be written
as a product of three Euler-like rotations, each described by one of the
mixing angles:

\begin{equation}
\rm{U}= 
\left(
\begin{array}{ccc}
1 & 0 & 0\\ 
0 & c_{23} & s_{23} \\  
0 & -s_{23} & c_{23} 
\end{array} \right)
\left(
\begin{array}{ccc}
c_{13} & 0 & s_{13}e^{i\delta}\\ 
0 & 1 & 0 \\  
-s_{13}e^{i\delta} & 0 & c_{13} 
\end{array} \right)
\left(
\begin{array}{ccc}
c_{12} & s_{12} & 0\\ 
-s_{12} & c_{12} & 0 \\  
0 & 0 & 1 
\end{array} \right)
\end{equation} 

where ``$s$'' represents sine of the mixing angle and ``$c$'' represents
cosine.

\begin{figure}[ht]
\centerline{\epsfxsize=3.9in\epsfbox{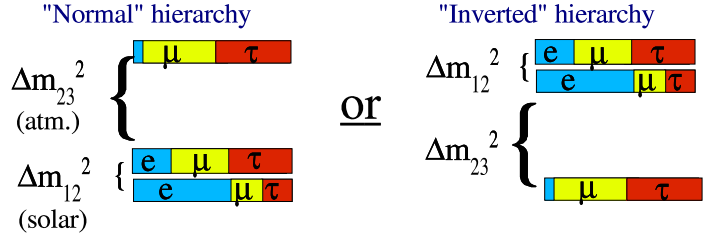}}   
\caption{Normal and inverted hierarchies in the three-flavor picture, 
with mass-squared values of the three states indicated vertically, and
possible flavor composition of the mass states indicated by the
horizontal divisions.\label{fig: numasses}}
\end{figure}

The ``1-2'' matrix describes solar mixing; the ``2-3'' matrix
describes atmospheric neutrino mixing.
The ``1-3'' or ``e3'' mixing is known to be small; $\theta_{13}$ may
be zero.
The mass-squared difference $\Delta m^2_{23}\sim 2 \times
10^{-3}$~eV$^2$ describes the atmospheric mixing, and 
$\Delta m^2_{12} \sim 4.5 \times 10^{-5}$ eV$^2$
describes solar mixing.  Neutrino oscillation
experiments tell us only about mass-squared differences;
the absolute mass scale is known only to be less than about 2~eV.
It is also as yet unknown whether the mass
hierachy is ``normal'', \textit{i.e.} the solar mixing is described
by  two lighter states, or ``inverted'', \textit{i.e.} the solar
mixing is described by two heavier states
(see Figure~\ref{fig: numasses}.)

The remaining questions can be addressed by neutrino oscillation
experiments are:

\begin{itemize}

\item Is U$_{e3}$ non-zero?
\item Is the hierarchy normal or inverted?
\item Is 2-3 mixing maximal, or just large?
\item Is the CP-violating phase non-zero?

\end{itemize}

\subsection{The Next Step: e3 Mixing}\label{theta13}

The next question which can be approached experimentally
is that of $e3$ mixing.  A consequence of a non-zero $U_{e3}$ matrix
element will be a small appearance of $\nu_e$ in beam of
$\nu_{\mu}$: for $\Delta m^2_{23}>>\Delta m^2_{12}$ (as is the
case), and for $E_{\nu} \sim L \Delta m^2_{23}$, ignoring
matter effects we find

\begin{equation}
P(\nu_{\mu}\rightarrow\nu_e)=\sin^22\theta_{13}\sin^2\theta_{23}\sin^2(1.27\Delta m_{23}^2 L/E).
\end{equation}

\begin{figure}[t]
\centerline
{\epsfxsize=8cm   
\epsfbox{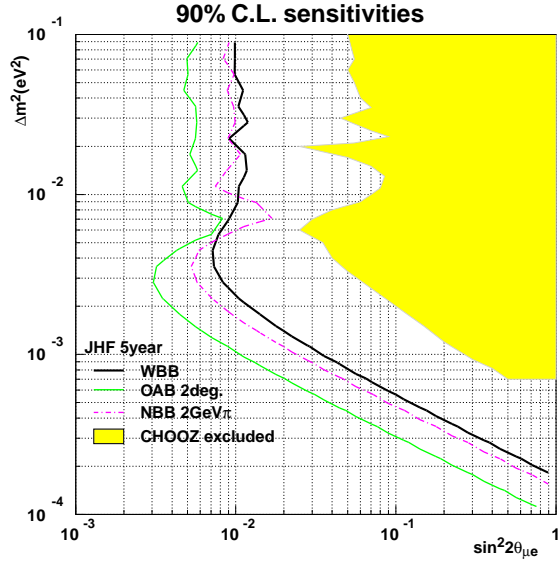}}
\caption{Expected sensitivity to $\sin^2 2\theta_{13}$ with J-PARCnu,
for various beam configurations (a 2$^\circ$ off-axis configuration,
labeled OAB 2deg on the plot, has been selected.)  The CHOOZ excluded
region for $\bar{\nu}_e$ disappearance is shown for comparison.  NuMi
off-axis and new reactor experiments have comparable
sensitivity.\label{fig:e3sens}}
\end{figure}

This expression illustrates that $\theta_{13}$  manifests itself
in the amplitude of an oscillation with 2-3-like parameters.
Since $\nu_e$ appearance
has never been observed at these parameters, 
this amplitude (and hence $\theta_{13}$) must
be \textit{small}.
The best limits so far, shown in
Figure~\ref{fig:e3sens}, come from a reactor experiment, CHOOZ, which
observed no disappearance of reactor $\bar{\nu}_e$.\footnote{In the literature one finds limits and sensitivities to this
mixing angle variously expressed in terms of $\theta_{13}$ (in radians
or degrees), $\sin\theta_{13}$, $\sin^2\theta_{13}$, $\sin^2
2\theta_{13}$, $\sin^2 2\theta_{\mu e}$, $|\rm{U}_{e3}|$, or
$|\rm{U}_{e3}|^2$; no convention has yet emerged.  For $\Delta
m_{12}^2 << \Delta m^2_{23} \sim \Delta m^2_{13}$, $\sin^2
2\theta_{\mu e} \equiv \sin^2 2\theta_{13} \sin^2 \theta_{23}$ is the
measured $\nu_e$ appearance amplitude. $|\rm{U}_{e3}|= \sin
\theta_{13}$, and for $\theta_{23} \sim \pi/4$ and $\theta_{13}$
small, it follows that $\sin^2 2\theta_{\mu e} \sim \frac{1}{2}\sin^2
2\theta_{13} \sim 2 \sin^2 \theta_{13}$.}.  The on-axis long baseline
experiments mentioned in section~\ref{longbaseline} can likely improve this
limit by a factor of approximately five.  To do better than this is a
difficult job: since the modulation may be parts per thousand or
smaller, one needs both good statistics and low background data.  The
primary sources of background for a long baseline experiment are:
intrinsic beam $\nu_e$ contamination, misidentified NC resonant
$\pi^0$ production (since $\pi^0$ decay to $\gamma$-rays which make
electron-like showers), and other misidentified particles.

\subsubsection{Off-Axis Beams}

\begin{figure}[ht]
\centerline
{\epsfxsize=8cm   
\epsfbox{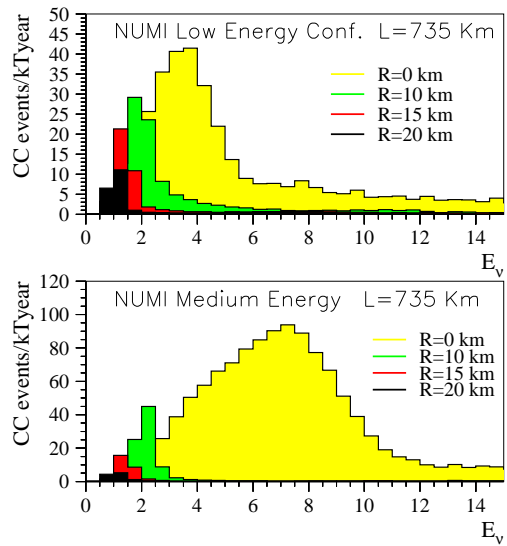}}
\caption{Expected CC interaction spectra for various
off-axis locations for low and medium-energy NuMI beams, from Reference~\protect\cite{pdriver}.}
\label{fig:offaxisspec}
\end{figure}

A promising next step for measurement of $\theta_{13}$ (assuming it is
large enough to be measured) is an off-axis detector at a long
baseline neutrino beam.  An experiment placed a few degrees off axis
has some kinematic advantages.  Because two-body pion decay kinematics
imply that neutrino energy becomes relatively independent of pion
energy off-axis, the neutrino spectrum becomes more sharply
peaked.  Therefore an off-axis siting is favorable for
background reduction and oscillation fits, in spite of the reduction
in $\nu$ flux.

There are two major long-baseline off-axis detector projects currently
under consideration.  The first of these is J-PARCnu\cite{jparc},
comprising a high power (0.77 MW) beam from the J-PARC facility in
Japan (currently under construction.)  The far detector, 2$^\circ$
off-axis, will be a fully refurbished Super-K at 295~km.  The second
off-axis proposal\cite{numioff}, for the U.S., is to exploit the NuMi
beam which will exist for MINOS, and build a new detector off-axis at
a distance of 700-900~km.  Various detector technologies and sites are
under discussion.  Sensitivity to $\theta_{13}$ for both possibilities
will be roughly a factor of 10-20 better than the CHOOZ limit.

\subsubsection{Reactor Experiments}

Another possibility currently under consideration by groups in Russia,
Japan and the U.S. is an upgraded reactor experiment employing the
same strategy as CHOOZ, i.e. search for disappearance of
$\bar{\nu}_e$\cite{reactor}.  The challenge for this type of
experiment is reduction of systematics to the level where a few
percent or smaller modulation is evident.  Multiple detectors may help
to achieve this.

\subsection{Leptonic CP Violation and Mass Hierarchy}

The long-term goal of long baseline oscillation experiments in
the observation of CP violation in the lepton sector.  The 
basic idea is to measure a difference
between $\nu_{\mu} \rightarrow \nu_e$  and
$\bar{\nu}_{\mu} \rightarrow \bar{\nu}_e$ transition
probabilities.  However, it is not simple to extract
a CP violating phase from the measurements: transition rates
depend on all MNS matrix parameters, and in addition
are affected by the presence of matter.

Following the analysis of Reference~\cite{cervera}: the 
approximate transition
probabilities for neutrinos and antineutrinos 
(assuming $\theta_{13}$, $\Delta_{12}L$ and
$\Delta_{12}/\Delta_{13}$ are all small) are:

\bea
P_{\nu_ e \nu_\mu ( \bar \nu_e \bar \nu_\mu ) } & = & 
s_{23}^2 \sin^2 2 \tetaot \, \left ( \frac{ \delot }{ \tilde B_\mp } \right )^2
   \, \sin^2 \left( \frac{ \tilde B_\mp \, L}{2} \right) \\
& + & 
c_{23}^2 \sin^2 2 \theta_{12} \, \left( \frac{ \Delta_{12} }{A} \right )^2 
   \, \sin^2 \left( \frac{A \, L}{2} \right ) \nn \\
& + & \label{approxprob}
\tilde J \; \frac{ \Delta_{12} }{A} \, \frac{ \delot }{ \tilde B_\mp } 
   \, \sin \left( \frac{ A L}{2}\right) 
   \, \sin \left( \frac{\tilde B_{\mp} L}{2}\right) 
   \, \cos \left( \pm \delta - \frac{ \delot \, L}{2} \right ) \nn \, , 
\eea

where 
\be
\tilde J \equiv c_{13} \, \sin 2 \theta_{12} \sin 2 \tetatt \sin 2 \tetaot,
\ee
$L$ is the baseline, $\Delta_{ij} \equiv \frac{\Delta m^2_{ij} }{2 E_\nu}$, $\tilde B_\mp \equiv |A \mp \delot|$, and
$A$ is the matter parameter 
$A=\sqrt{2} G_F N_e$, where $G_F$ is the Fermi constant and $N_e$ is
the electron density of the matter traversed.  The upper sign in $\pm$
refers to neutrinos and the lower to antineutrinos.
The first two terms are the ``non-CP'' terms
and the last term depends on the CP-violating phase.

A few observations about observability
of leptonic CP violation may be made based on this expression:

\begin{itemize}
\item The fact that $\sin^2 2\theta_{12}$ is large, as recently 
indicated by SNO and KamLAND, is good news: the CP term is
proportional to $\sin 2\theta_{12}$.

\item The CP terms are proportional to $\sin 2\theta_{13}$; therefore
a very small value of $\theta_{13}$ will mean that
CP violation will be very difficult to observe.

\item Precision measurements of all parameters will be necessary
for measurement of CP observables!

\item Matter effects cause a fake asymmetry between neutrinos and
antineutrinos, as indicated by the matter terms (see 
Figure~\ref{fig:nunubar} from Reference~\cite{pdriver}.)  This may be
considered a blessing rather than a curse, because the sign of the
neutrino/antineutrino asymmetry depends on the sign of $\Delta m^2$.
In other words, one can learn about the mass hierarchy by measuring
the asymmetry.  Observation of matter effects requires relatively long
baselines (typically more than 500~km, depending on parameters.)

\item Parameter ambiguities from various sources
are intrinsic to these equations, and a single measurement will not
suffice to measure both $\delta$ and $\theta_{13}$ (and matter
effects.)  For instance, consider Figure~\ref{fig:minakataplot} drawn
from Reference~\cite{minakata2}, which
shows probability of transition for antineutrinos versus probability
of transition for neutrinos (the axes represent the observables.)  
For a given value of $\sin^2 2\theta_{13}$,
one can draw an ellipse corresponding to different values of $\delta$;
this ellipse is shifted for different hierarchies via the matter effect.
In consequence, one must make multiple measurements with different
experimental parameters in order to resolve the ambiguities.
Some of these issues are explored in \textit{e.g.} References~\cite{minakata2},\cite{ambiguity}.

\begin{figure}[!htb]
\centerline
{\epsfxsize=6cm   
\epsfbox{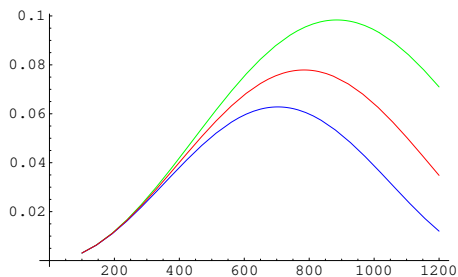}}
\caption{Transition probabilities for neutrinos (green, top curve)
and anti-neutrinos
(blue, bottom curve) in matter and vacuum  (red, middle curve)
as function of the distance for 2~GeV,  $\Delta m^2_{13}=
3\times 10^{-3}~{\mbox{eV}}^2$ (normal hierarchy),
$\theta_{\rm 23}=\pi/4$,
$\Delta m^2_{12} = 1\times 10^{-4}$~eV$^2$, $\theta_{23}=\pi/6$, $|U_{e3}|^2=0.04$,
and $\delta=0$. Figure from Reference~\protect\cite{pdriver}.}
\label{fig:nunubar}
\end{figure}

\begin{figure}[ht]
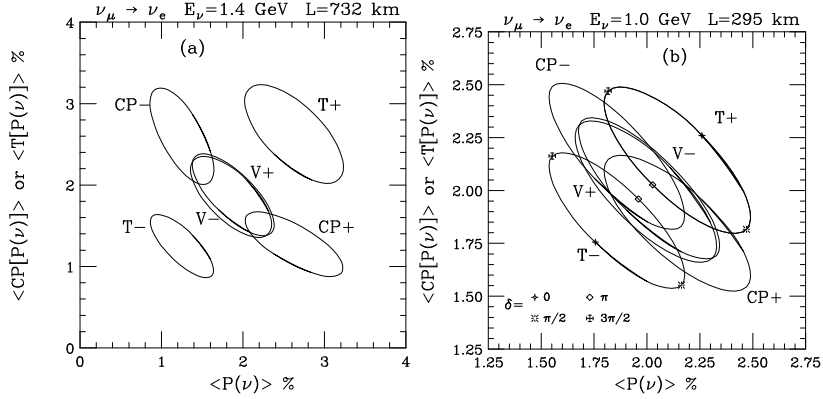

\begin{minipage}[t]{2.1in}
\begin{centering}
\epsfxsize=2.1in \epsfbox{Fig4a_mnp.epsi}
\end{centering}
\end{minipage}
\begin{minipage}[t]{2.1in}
\begin{centering}
\epsfxsize=2.1in \epsfbox{Fig4b_mnp.epsi}
\end{centering}
\end{minipage}
\caption{The T (CP) trajectory diagrams (ellipses) in the plane
$P(\nu_\mu \rightarrow \nu_e)$ versus $P(\nu_e \rightarrow \nu_\mu)$
($P(\bar{\nu}_\mu \rightarrow \bar{\nu}_e)$) for an average
neutrino energy (spread 20\%) and baseline of (a) 1.4 GeV and 732 km
(NUMI/MINOS) and (b) 1.0 GeV and 295 km (J-PARCnu.)
The ellipses labeled with a T and CP are in matter with
a density times electron fraction given by $ Y_e \rho  = 1.5$ g cm$^{-3}$
whereas those ellipses labeled V are in vacuum where the T and CP
trajectories are identical.
The plus or minus indicates the sign of $\Delta m^2_{31}$.
The mixing parameters are fixed to be
$|\Delta m^2_{31}| = 3 \times 10^{-3}$ eV$^2$,
$\sin^2 2\theta_{23}=1.0$,
$\Delta m^2_{21} = +5 \times 10^{-5}$ eV$^2$,
$\sin^2 2\theta_{12}=0.8$ and
$\sin^2 2\theta_{13}=0.05$.
The marks on the CP and T ellipses are the points where
the CP or T violating phase
$\delta = (0,1,2,3){\pi/2}$ as indicated. See Reference~\protect\cite{minakata2} for details.}\label{fig:minakataplot}
\end{figure}

\end{itemize}

Assuming that $\theta_{13}$ is large enough for there to be some
hope of observing leptonic CP violation, the obvious strategy is
an upgraded long baseline experiment, perhaps as a ``Phase II'' of
an off-axis program, or perhaps
as an on-axis detector program such as the proposed
broad-band beam from Brookhaven\cite{diwan}.  
Very large detectors (e.g. the 1~Mton Hyper-K detector)
and upgraded ``superbeams'' have been proposed\cite{jparc,diwan,uno,nusl}.  As mentioned above,
multiple measurements with differing energies and/or baselines,
with both neutrinos and antineutrinos will be necessary for
full characterization of the parameters.

More ambitious alternatives to a traditional proton-induced
neutrino superbeam have been proposed.
For instance a muon storage ring ``neutrino factory'' would
produce copious, and well-understood, 20-50~GeV neutrinos from muon
decay\cite{nufact}.  Detectors at 3000-7000~km could explore
matter effects.  However this idea is for the rather distant
future due to high cost and technical difficulties to be overcome.
Other interesting ideas include a ``beta-beam'' of radioactive
ions which could provide a high flux of $\bar{\nu}_e$'s\cite{betabeam,ellis}.

\section{Non-Oscillation Neutrino Physics}

So far this review has focused on neutrino oscillation studies.  However,
oscillation physics hardly comprises all of
neutrino physics.  Perhaps the two most compelling experimental
questions that cannot be answered by oscillation experiments
are: 

\begin{itemize}
\item What is the absolute mass scale? 
We do not know whether the masses are hierarchical or 
degenerate. This question is fundamental,
and additionally has profound consequences for cosmology\cite{ellis}.

\item Are neutrinos Majorana or Dirac?  In other words, are they
their own antiparticles, described by a two-component spinor,
or described by a 4-component Weyl spinor?  The answer to this
question has tremendous implications for the construction of theory
describing neutrino masses.  For instance, the ``see-saw'' mechanisms
for neutrino mass generation~\cite{bnv} require the neutrino to be Majorana.
\end{itemize}

In the following sections I will very briefly review experiments which
aim to answer some, or both, of these questions.  In lieu of
detailed discussion, I will point to comprehensive reviews
where possible.

\subsection{Kinematic Neutrino Mass Experiments} 

As noted above, neutrino oscillation measurements say nothing about
absolute masses of the mass states.  The idea behind kinematic
neutrino mass searches is simple: look for missing energy.  The
traditional tritium beta decay spectrum endpoint experiments now have
limits for absolute $\bar{\nu}_e$ mass from the Mainz and Troitsk
experiments of $\lsim$3~eV\cite{pdb}, and there are some prospects for
improvement down to the sub-eV level by the Katrin\cite{katrin}
experiment.  Some new techniques are under consideration,
too\cite{newkinem}.  The $\nu_{\mu}$ and $\nu_{\tau}$ mass limits are
currently 190~keV\cite{numukin} and 15.5~MeV\cite{nutaukin}
respectively; however improving these direct $\nu_\mu$ and $\nu_\tau$
measurements seems less compelling if information about differences
between the mass states is available from oscillation experiments.

\subsection{Double Beta Decay} 

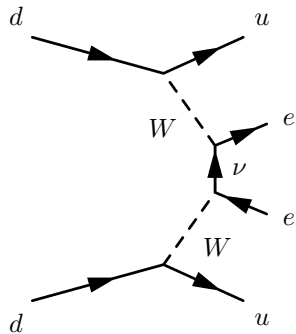
\begin{figure}[ht]
\begin{center}
\begin{fmffile}{three}
  \fmfframe(1,7)(1,7){ 
   \begin{fmfgraph*}(100,100)
	\fmfleftn{i}{2}
	\fmfrightn{o}{4}
	 \fmflabel{$d$}{i1}
	 \fmflabel{$d$}{i2}
	 \fmflabel{$u$}{o1}
	 \fmflabel{$u$}{o4}
	 \fmflabel{$e$}{o2}
	 \fmflabel{$e$}{o3}
         \fmf{fermion}{i1,v1,o1}
         \fmf{fermion}{i2,v4,o4}
         \fmf{dashes,label=$W$}{v1,v2}
         \fmf{dashes,label=$W$}{v3,v4}
         \fmf{plain,label=$\nu$}{v2,v3}
         \fmf{fermion}{o2,v2,v3,o3}
   \end{fmfgraph*}
  }
\end{fmffile}

\caption{Underlying process of neutrino-less double beta decay, which
requires the neutrino to be its own antiparticle.\label{fig:0nubbdk}}
\end{center}
\end{figure}

Another way of getting at absolute neutrino mass, and, in one fell
swoop, determine that the neutrino is Majorana, is to discover
neutrinoless double beta decay, $(N,Z)
\rightarrow (N-2,Z+2)+e^-+e^-$.  Such a decay is only possible if the
neutrino has mass, and is Majorana, as illustrated in
Figure~\ref{fig:0nubbdk}.  The current 90\% confidence level lowest
mass limits from non-observation of double beta decay are $\langle
m_{\nu} \rangle = | \Sigma{U_{1j}^2 m_{\nu j}}| < 0.35$ eV\cite{pdb}
(note dependence of these limits on the matrix elements.) The current
best limits are from $^{76}$Ge experiments.  Many new double beta decay
search experiments are planned and under construction, 
some employing novel techniques.  It appears challenging
but not impossible to push
the limits down to
$\sim$0.02~eV\cite{bbdkreview}.

\subsection{Neutrino Magnetic Moment}
A non-zero non-transition magnetic moment would imply that neutrinos
are Dirac and not Majorana.  The best limits on neutrino magnetic
moment\cite{pdb}, in the range $10^{-12}\mu_B$, are astrophysical.
The current best laboratory limit, $\mu_{\nu_e}<1.0 \times
10^{-10}\mu_B$, is from the MUNU experiment\cite{munu}, which measures
low energy elastic scattering of reactor $\bar{\nu}_e$ on electrons.

\subsection{Supernova Neutrinos}
A supernova is a ``source of opportunity'' for neutrino physics: we
can expect an enormous burst of neutrinos of all flavors from a core
collapse in our Galaxy about once every 30~years.  Many of the large
neutrino experiments -- Super-K, SNO, Borexino, KamLAND, LVD and
AMANDA, and BooNE -- are sensitive\cite{snnu} to a burst of supernova
neutrinos.  Most of these are water or
scintillator-based and are primarily sensitive to $\bar{\nu}_e$.  In
addition to bringing new understanding of stellar core collapse
processes, a Galactic supernova would be a tremendous opportunity
for neutrino physics.  We may be able to extract absolute mass
information from time-of-flight-related measurements, although such
information, in the best case, may be only marginally better than the
best kinematic (and cosmological) limits\cite{snkinem}.  Potentially
more promising is the information about mixing parameters
($\theta_{13}$ and mass hierarchy) that may be inferred from spectra
and time-dependence of the different flavor components of the
flux\cite{snosc}: matter-induced flavor transitions in the stellar
material, and also in the Earth, may cause inversion of the expected
hierarchy of temperatures
($\bar{E}_{\nu_{\mu},\nu_{\tau},\bar{\nu}_{\mu},\bar{\nu}_{\tau}} >
\bar{E}_{\bar{\nu}_e}> \bar{E}_{\nu_e}$) for some mixing parameters
and conditions.  Therefore detectors with sensitivity to flavors other
than $\bar{\nu}_e$ are highly desirable.

The relic supernova neutrinos  (neutrinos from past supernovae)
have not yet been observed; best limits so far come from 
Super-Kamiokande\cite{snrelic}.

\subsection{Cosmology} Neutrinos are important
in cosmology\cite{cosmology}, and play an significant role in big-bang
nucleosynthesis and possibly ``leptogenesis'', the process
by which the matter-antimatter asymmetry of the universe
was generated.  Ultra low energy (1.95~K)
big-bang relic neutrinos are expected to permeate the universe with a
number density of 113 cm$^{-3}$ per family.  The relic neutrinos must
make up some component of the dark matter, although the fraction is now
thought to be small, for consistency with galactic structure
formation. Direct detection of these big bang relic neutrinos remains
an experimental challenge\cite{relic}.  However, recent advances in
``precision cosmology''\cite{ellis,pogosyan} are starting to provide
quite strong constraints on the properties of neutrinos.  The latest
cosmic microwave background (CMB) anisotropy data from the WMAP
satellite, combined with other CMB experiment data, and other data
such as the 2dF Galaxy Redshift Survey, constrain the sum of absolute
masses of the neutrino states to be less than $\sim$2 eV (depending
on assumptions)\cite{cosmology2}. These results now rival laboratory
kinematic limits, and constrain scenarios involving sterile
neutrinos\cite{cosm_constr}. There are even prospects for attaining
sensitivity to neutrino masses as low as 0.1~eV via precision
cosmological measurements (\textit{e.g} Reference~\cite{Abazajian}.)

\subsection{Neutrino Astrophysics}
Very large area long string water Cherenkov detectors (AMANDA, Baikal,
Antares, Nestor and the next-generation kilometer-scale IceCube) in
ice and water are embarking on a new era of high energy neutrino
astronomy\cite{longstring}.  ``Cosmic particle accelerators'' such as
active galactic nuclei and the cataclysmic events that produce
gamma-ray bursters, are expected to produce $\sim$ PeV neutrinos, visible in
these detectors as upward-going events.  Exotic astrophysical sources
that produce ultra-high-energy neutrinos may be associated with the
highest energy cosmic rays, too.  In fact, detectors designed
primarily to explore ultra-high-energy cosmic ray air showers will
have sensitivity to neutrino-induced ``earth-skimming'' horizontal air
showers\cite{horiz_shower}.  Because the Earth starts to become
opaque, via CC interactions, to $\nu_{\mu}$ and $\nu_e$ at a few
hundred TeV, and $\nu_{\tau}$'s ``regenerate'' as the CC-induced
$\tau$'s decay, the flavor content of the flux (and the effect of
oscillations) can be explored by looking at the angular distribution
of observed events.  From these neutrinos, we may learn about the
physical mechanisms behind the ultra-high-energy sources.  Measurement
of the relative timing between neutrinos and photons from the same
source will test special relativity and the weak equivalence
principle.

\section{Summary}

It is now quite certain that neutrinos have mass and mix: atmospheric
and solar oscillation signals are now multiply confirmed
and parameters are quite well constrained.  The LSND
observation does not fit in to the three-flavor picture; we await
BooNE to confirm or refute it.  If BooNE confirms the LSND appearance
observation, we will have to begin exploring the possibilities
required to explain it.  If not, and the standard three-flavor
scenario holds, the next steps for neutrino oscillation experiments
are clear: search for non-zero $\theta_{13}$, determine the mass
hierarchy and ultimately, if parameters are favorable, go for leptonic
CP violation.  On the non-oscillation front, absolute mass can be
approached via kinematic experiments and double beta decay, the latter
also promising insight into the Majorana-vs-Dirac question.  Precision
cosmology is also making headway towards understanding of
neutrinos. Knowledge of these parameters is essential for full
understanding of matter-antimatter asymmetry of the universe, as well
as a full description of the fundamental particles and their
interactions. Although this last decade will be a hard act to follow
for excitement in neutrino physics, the next decade promises yet more
entertainment.

\end{document}